# Epitaxial Growth of Two-dimensional Insulator Monolayer Honeycomb BeO

AUTHOR NAMES Hui Zhang,[†,‡,#] Madisen Holbrook,[†,#] Fei Cheng,[†] Hyoungdo Nam,[†] Mengke Liu,[†] Chi-Ruei Pan,[§] Damien West,[‖] Shengbai Zhang,[‖] Mei-Yin Chou,[§,⊥] and Chih-Kang Shih[†,*]

[†]Department of Physics, The University of Texas at Austin, Austin, TX 78712, USA

[§]School of Physics, Georgia Institute of Technology, Atlanta, Georgia 30332, USA

[‖]Department of Physics, Rensselaer Polytechnic Institute, Troy, New York 12180, USA

[⊥]Institute of Atomic and Molecular Sciences, Academia Sinica, Taipei 10617, Taiwan

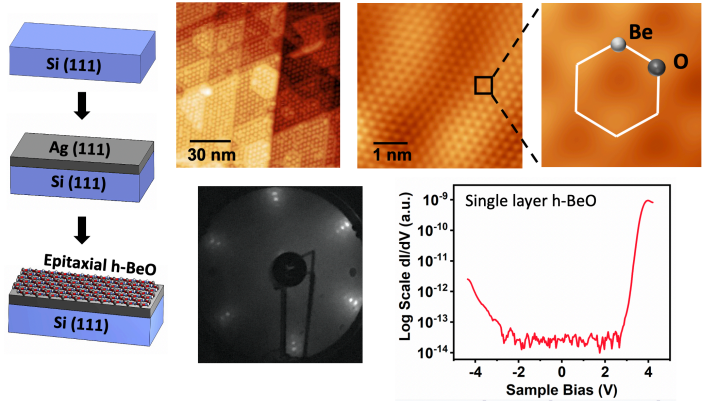

The emergence of two-dimensional (2D) materials launched a fascinating frontier of flatland electronics. Most crystalline atomic layer materials are based on layered van der Waals materials with weak interlayer bonding, which naturally leads to thermodynamically stable monolayers. We report the synthesis of a 2D insulator comprised of a single atomic sheet of honeycomb structure BeO (h-BeO), although its bulk counterpart has a wurtzite structure. The h-BeO is grown by molecular beam epitaxy (MBE) on Ag(111) thin films that are conveniently grown on Si(111) wafers. Using scanning tunneling microscopy and spectroscopy (STM/S), the honeycomb BeO lattice constant is determined to be 2.65 Å with an insulating band gap of 6 eV. Our low energy electron diffraction (LEED) measurements indicate that the h-BeO forms a continuous layer with good crystallinity at the millimeter scale. Moiré pattern analysis shows the BeO honeycomb structure maintains long range phase coherence in atomic registry even across Ag steps. We find that the interaction between the h-BeO layer and the Ag(111) substrate is weak by using STS and complimentary density functional theory calculations. We not only demonstrate the feasibility of growing h-BeO monolayers by MBE, but also illustrate that the large-scale growth, weak substrate interactions, and long-range crystallinity make h-BeO an attractive candidate for future technological applications. More significantly, the ability to create a stable single crystalline atomic sheet without a bulk layered counterpart is an intriguing approach to tailoring novel 2D electronic materials.

**Introduction**

The discovery of graphene and its unprecedented properties inspired an extraordinary increase in research progress, launching an era of two-dimensional (2D) electronic materials.[1-7] One property these emerging materials share is strong in-plane bonding with weak interlayer interactions, leading to the existence of stable crystalline atomic layers. This enables the design of ultrathin 2D devices with different crystalline atomic layers as the building blocks. While many conceptual devices are fabricated by mechanical exfoliation, in order to create a scalable technology platform, it is critical to develop an epitaxial growth of single crystalline 2D materials. Early exploration of 2D materials focused on van der Waals (vdW) crystals, which have a naturally occurring bulk layered structure, and have already demonstrated significant potential for 2D technological applications. Synthesis of epitaxial wafer scale monolayers has been achieved for graphene[8] and hBN,[9] proving that large scale growth is feasible for vdW materials. While the success of vdW crystals is promising, the family of layered bulk solids is limited, constraining the variety of structural and electronic properties available for device design.

Driven by the desire to expand the library of 2D materials, there has been increased exploration into the synthesis of stable atomically thin phases of electronic materials without a bulk layered structure. For example, 2D allotropes of group IV elements silicene, germanene, and stanene have been successfully grown and are predicted to have attractive electronic properties.[10-14] However, unlike graphene, these elements do not favor sp2 hybridization and adopt a mixed sp2-sp3 hybridization, resulting in a buckled honeycomb structure. This lack of a stable planar phase creates challenges such as multi-phase growth, buckled superstructures, and strong substrate interactions.[10-14] By contrast, polar wurtzite structure materials were surprisingly predicted to circumvent the bulk thermodynamic constraint at the few layer limit, resulting in a stable planar sp2 hybridized graphitic phase.[15] In the thin regime of a few atomic layers, polar ionic bonds may turn into the in-plane direction, forming a honeycomb structure to diminish unstable surface polarization (shown schematically in Figure 1a).[15, 16] This theory indicates that initial growth forms a graphitic film, and at a critical layer thickness, the crystal becomes unstable and converts to a wurtzite structure.[15] These predictions have stimulated intensive experimental investigations of the graphitic phases of bulk wurtzite structure materials,[17, 18] with ZnO grown on (Ag, Au) substrates drawing the most attention.[19-21] Motivated by the success of these experimental studies, the question naturally arises as to whether other wurtzite structure materials can be epitaxially grown as a 2D graphitic layer.

Beryllium oxide (BeO) is a polar wurtzite structure material with remarkable physical and mechanical properties, such as a high thermal conductivity, hardness, and a large insulating bandgap of 10.7 eV. These attractive properties have inspired significant theoretical research interest concerning the existence of other possible structural phases of BeO. In particular, as a member of the isoelectronic series of first row elements like graphene and hBN, studies predicted BeO might also exist in a $sp^2$ hybridized atomic layer with a honeycomb structure (h-BeO).[22, 23] Later, monolayer graphitic h-BeO was predicted to be energetically stable in free space, and that it could be synthesized from BeO molecules.[24] Further, not only was h-BeO predicted to have the most stable graphitic phase of wurtzite materials,[15] but theoretical surveys of possible monolayer honeycomb structures of II-VI semiconductors predicted h-BeO has the lowest formation energy corresponding to the bulk phase,[25] and was deemed the most experimentally achievable.[16] Though h-BeO is predicted to exist at the ultra-thin limit, synthesis of this intriguing material has remained elusive until recently. The only experimental evidence of planar h-BeO to date was achieved by a wet chemistry approach in a graphene liquid cell, but the growth was ~20 layers thick and polycrystalline on the order of tens of nanometers.[26] Nevertheless, these promising theoretical and experimental results strongly suggest that the epitaxial growth of crystalline, monolayer h-BeO is experimentally achievable.

In this work, we report the realization of the epitaxial growth of a single atomic sheet of h-BeO grown on Ag(111) substrates using molecular beam epitaxy (MBE). Here, we employ scanning tunneling microscopy (STM) to directly characterize the atomic and electronic structure of the BeO, and further show that the monolayers maintain high crystallinity across the Ag terraces. Using low energy electron diffraction (LEED), we reveal the h-BeO nanosheet covers a nanometer scale wafer and maintains uniform crystallinity at this length scale. Finally, using STS and complimentary density functional theory (DFT) calculations, we show that the h-BeO weakly interacts with the Ag surface.

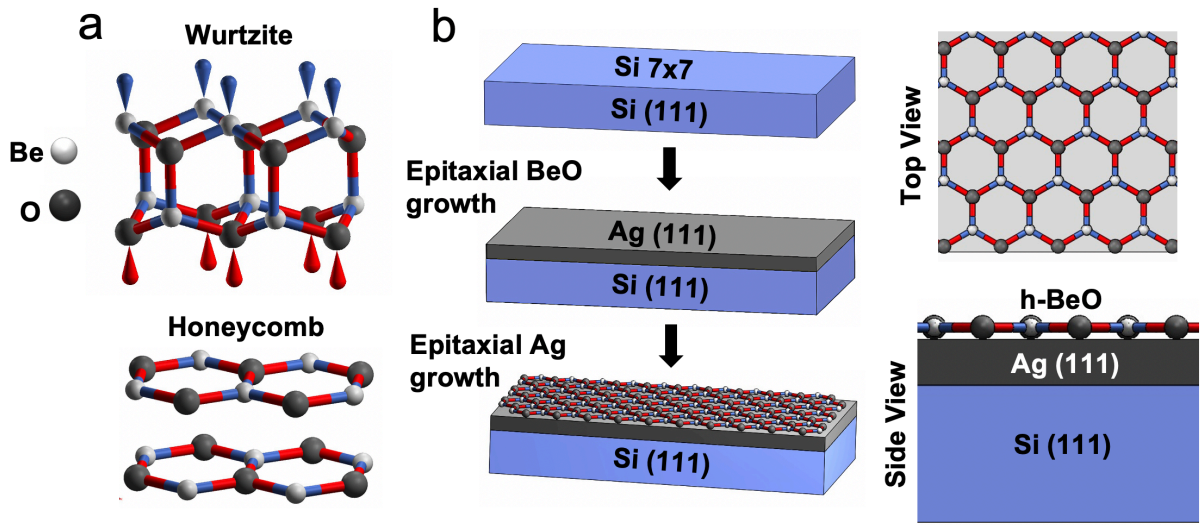

**Figure 1.** (a) Schematic of the bulk wurtzite (top) and layered graphitic honeycomb (bottom) phases of BeO. The Be and O atoms are white and grey, respectively. (b) Schematic of the growth

process of the honeycomb BeO. From top to bottom, a Si(111) substrate is prepared with a 7x7 reconstruction, a 150 nm Ag film is epitaxially grown, and honeycomb BeO is grown epitaxially on top. To the right is a top and side view of the heterostructure.

**Results and Discussion**

Honeycomb structure BeO was epitaxially grown on Ag(111) substrates using the growth process shown schematically in Figure 1b. First, a 150 nm thick Ag(111) thin film template is grown by MBE on a Si(111) wafer substrate with a 7x7 surface reconstruction at room temperature. Next, the h-BeO is grown epitaxially on top of the Ag thin film by MBE, with the Ag substrate held at 500 °C, and a Knudsen cell used as an evaporation source (See Methods for a detailed growth summary). This two-step epitaxial growth method on Si substrates, which are widely used in industry, makes the h-BeO growth conveniently scalable.

The initial growth of the h-BeO was studied by applying a submonolayer coverage of 0.4 monolayer, presented in STM images at different bias conditions, $V_s = 4.1$ V and 1 V, respectively (Figure 2a and 2b). As discussed below, the h-BeO has a relatively large bandgap with the conduction band minimum (CBM) located at ~ $3.0 \pm 0.2$ eV above $E_F$. Consequently, Figure 2a (above the CBM) shows the formation of well-defined h-BeO islands that can be identified by a distinct moiré pattern, while in Figure 2b (below the CBM) the h-BeO regions, outlined in black for clarity, are almost indistinguishable from the Ag surface. When the sample bias is within the h-BeO bandgap, quantum tunneling occurs only from the Ag states through the BeO layer, resulting in an image that reflects the structure of the underlying Ag (Figure 2b). Although the surface exhibits a multi-step morphology, a careful comparison of the bias dependent images allows us to conclude these are single layer h-BeO 2D islands on Ag(111). Figure 2c shows line profiles corresponding to the black and blue lines in Figures 2a and 2b, with the arrow denoting the direction of the line cut. The thickness of the h-BeO is determined by comparing the black line cuts across the h-BeO island shown in the top graph of Figure 2c. At 1 V (solid line), the profile is nearly flat except for a small depression of 0.3 Å at the BeO layer due to a shift in the Ag work function (discussed later). At 4.1 V, the BeO step height appears ~3.2 Å above the Ag substrate (see Figure 4c), which is consistent with the predicted interlayer spacing of the hypothetical graphitic BeO structure,[22] as well as our theoretical calculation (discussed below). The blue line cuts in Figures 2a and 2b cross a feature that resembles a second layer of h-BeO at first sight, but the corresponding line profiles (bottom graph of Figure 2c) validate the existence of a single h-BeO layer. We determine a step height of ~2.2 Å (green arrow) which correlates to the underlying Ag terrace height, while the difference in height of the two curves (red arrows) clarifies that h-BeO covers the Ag surface in a uniform monolayer.

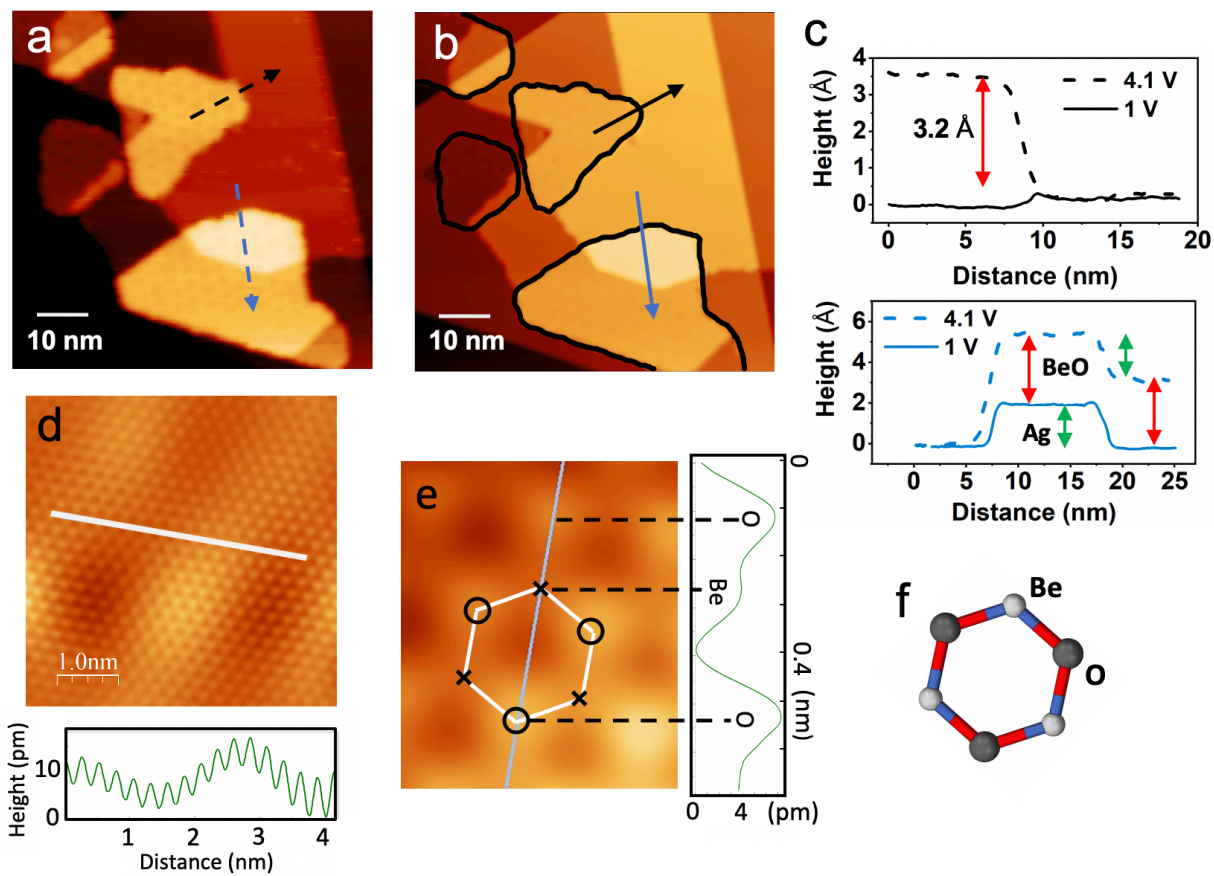

**Figure 2.** STM images of a 0.4 ML growth of h-BeO acquired at $I_t = 5$ pA and (a) 4.1 V and (b) 1V sample bias (area: 70 x 70 nm²). The black and blue arrows correspond to the line profiles in (c) showing the h-BeO islands are monolayers with a height of ~ 3.2 Å. (d) Atomic resolution STM image of BeO (5 x 5 nm²), $I_t = 30$ pA, $V_s = 50$ mV. Below is a line profile corresponding to the white line showing the atomic corrugation (e) Zoom in view of a honeycomb unit cell showing the different atomic sites of the Be and O. To the right a line cut across the honeycomb shows the profile of the atomic sites with (f) showing a corresponding schematic of the BeO honeycomb.

High-resolution STM images acquired at $V_s = 0.05$ V (Figure 2d) verify that the BeO has a honeycomb atomic structure by direct observation. Below the atomic resolution image, a line profile shows the atomic corrugation of the h-BeO, which yields a lattice constant of $2.61 \pm 0.08$ Å. An enlarged view of the honeycomb structure is visible in Figure 2e, which shows the atomic lattice with the location of the atomic sites of the Be (cross) and O (circle) atoms marked within the unit cell, with a corresponding model shown in Figure 2f. The line profile along the diagonal direction of the unit cell reveals the location of the atoms, with the O atom at the primary maximum and the Be atom at the shoulder.

A full monolayer growth of h-BeO was further characterized by STM, as shown in Figure 3a-c. A 120 nm x 120 nm STM image (Figure 3a) of the Ag(111) thin film substrate shows triangular atomic terraces bordered by atomic steps (step height of $2.36 \pm 0.1$ Å). Figure 3b and 3c are STM images of a surface after a complete atomic sheet of honeycomb BeO was deposited, acquired at a sample bias of 2.5 V and 3.55 V, respectively. As discussed above, STM images obtained below the h-BeO CBM reflect the morphology of the underlying Ag surface, which explains the similarity of Figure 3b to the pristine Ag(111) in Figure 3a. The measured step heights in images 3b and 3c are $2.36 \pm 0.1$ Å, which matches the Ag terrace height, confirming the sample surface is covered with a complete monolayer of h-BeO. At a sample bias of 3.55 V (above the CBM), one observes a clear moiré pattern due to the beating between the atomic lattices of the h-BeO and Ag(111) surface (Figure 3c). We note that the contrast of the moiré pattern is curiously reversed from the pattern shown in Figure 2a, which will be clarified later. The hexagonal moiré pattern has a periodicity of 3.2 ± 0.1 nm and is rotationally aligned with the Ag substrate step along the ⟨110⟩ direction, which indicates that the Ag(111) and h-BeO lattices are also rotationally aligned.[27] Briefly, moiré pattern periodicity depends on the rotational misalignment and/or difference in lattice periodicity of the overlayer with respect to the substrate.[27] With this knowledge, we can exploit the moiré pattern to determine the h-BeO lattice constant with greater accuracy: with a moiré periodicity of 3.2 nm, and the known lattice constant of 2.89 Å for Ag (111), one can deduce a lattice constant of 2.65 ± 0.01 Å for the h-BeO. This calculated lattice constant is consistent with the atomic resolution measurements, displaying a reduction in BeO bond length compared to bulk wurtzite structure, as theoretically predicted.[16, 22, 24, 28-32] The decrease in lattice constant was previously theorized to stem from the graphitic Be-O bond having an increased covalent character, and as a result the atoms move closer to increase the orbital overlap of nearest neighbors.[16, 23, 28]

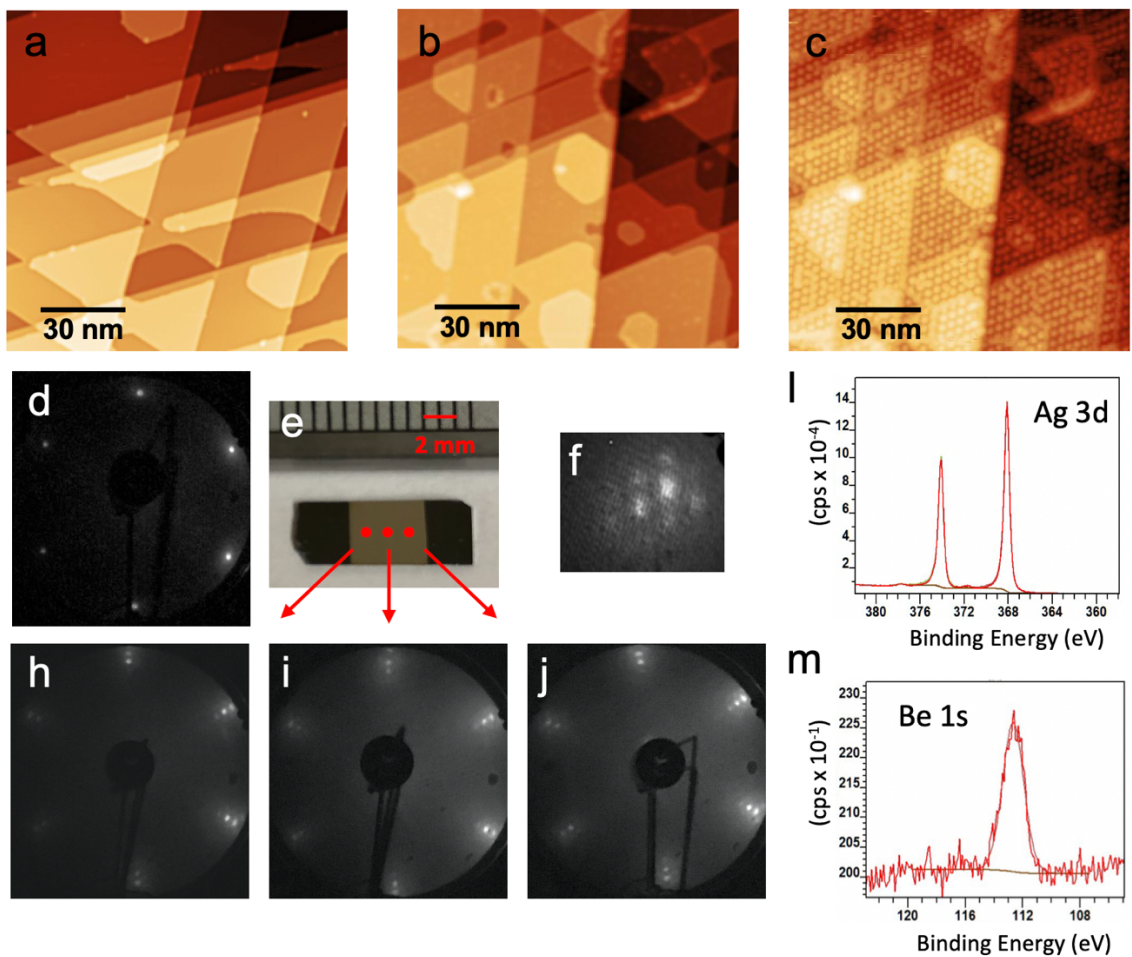

**Figure 3.** (a) STM image of an Ag (111) film, and (b, c) full coverage monolayer h-BeO surface, (area: 120 x 120 nm²) $I_t$ = 5 pA, $V_s$ = 2 V, 2.5 V, and 3.55 V respectively. (d) LEED pattern from Ag(111) surface acquired at 50 V electron beam voltage. (e) An image of the h-BeO sample on the silicon wafer, where the lighter area in the center is the location of the Si/Ag/BeO heterostructure, ~5 x 5 mm². The red dots indicate the locations of the LEED patterns (h, i, j) acquired with an electron beam energy of 58 V. (c) A zoom-in of the LEED pattern from the rotationally aligned Ag and BeO. (l, m) XPS measurements of the Ag 3d and Be 1s core levels. The Be/Ag atomic fraction ratio is determined to be 1:4.23.

In Figure 3d-j we utilize LEED to further examine the structure of the h-BeO film. Figure 3d shows the sharp hexagonal LEED pattern of the bare Ag(111) substrate, and after the growth of a BeO monolayer, additional spots appear in the pattern. In order to establish that the h-BeO forms a continuous layer, LEED patterns were obtained at different locations along the sample surface, indicated by the red dots in Figure 3e. Figures 3h-j show the LEED patterns of the full coverage h-BeO growth, providing additional evidence that the monolayer has a hexagonal structure rotationally aligned with the Ag(111) surface. A zoom in of the LEED pattern shows the formation of the moiré pattern (see Figure 3f). Each LEED measurement was obtained at a location more than one millimeter apart, reinforcing our assertion that the formation of the h-BeO is uniform across the wafer, similar to the wafer scale growth of graphene[8] and hBN.[9] As additional verification that the surface is BeO, we carried out *ex- situ* XPS (x-ray photoelectron spectroscopy) measurements of the Ag and Be core levels (Figure 4l, m), from which we determine a Be/Ag atomic ratio consistent with a single monolayer of honeycomb BeO on Ag(111).

Interestingly, not only do we find that the single atomic sheet can be extended across the whole wafer, but the phase of the atomic registry is also well maintained even across the Ag step. This conclusion is based on analysis of the moiré pattern; small changes in the lattice constants due to defects and strain are greatly magnified by the moiré pattern, making it an excellent tool to gauge the crystallinity of materials.[33] Shown in Figure 4a is a topographic image, and 4b the corresponding conductivity image, with the underlying Ag steps marked as *N*, *N+1*, *N+2*, and *N+3*. The conductivity image (Figure 4b) eliminates the height difference and planarizes the moiré patterns on different steps, improving the visibility of the pattern. The image is marked with lines to guide the eye to the rows of the moiré patterns at different levels, and across each step one observes an abrupt phase shift (labeled by δ) in the moiré pattern rows. This phase shift δ is roughly 1/3 of the spacing between the rows (labeled by λ), and the same phase shift is observed across each step, as shown by the model in Figure 4c. The underlying Ag atomic steps follow an ABC stacking sequence, resulting in a phase shift of 1/3 between the atomic rows in layer A and layer B (and between B and C) as shown by the stacking model in Figure 4d. Note that as one follows the blue dashed line across three Ag terraces from *N* to *N+3*, the moiré pattern rows return to their original alignment. This reveals that the BeO honeycomb lattice preserves the original phase as it continuously rolls across each step, with the shift in the Ag atomic rows creating a phase shift of $\frac{\delta}{\lambda} = \frac{1}{3}$ in the moiré pattern rows, as illustrated by the models (Figure 4c,d).

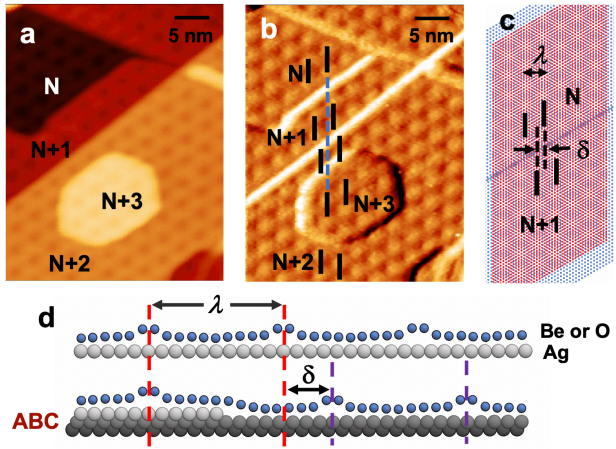

**Figure 4.** (a) A topographic STM image (40 x 40 nm²) of the surface of one monolayer of h-BeO, $I_t = 5$ pA, $V_s = 4.5$ V and (b) the corresponding conductivity image taken simultaneously by using lock-in. The images show a phase shift of $\delta = \lambda/3$ across the Ag steps, where $\lambda$ is the spacing of the moiré pattern rows, as indicated by the lines. c) A model showing that the atomic registry of the BeO is preserved with a phase shift of the moiré pattern due to the ABC stacking of the Ag(111) steps. d) A model showing a side view of the BeO layer rolling over the Ag steps. The moiré periodicity contains 12 BeO/11Ag(111).

The discovery of the high crystallinity of the h-BeO maintained even across Ag steps is fascinating. The low coverage (0.4 ML) scans in Figures 2a and 2b suggest the nucleation of h-BeO 2D islands occurs in the middle of terraces with an average island-island distance of ~ 50 nm. Combined with the above analysis, one can conclude that after nucleation in the middle of terraces, the 2D islands grow laterally over the Ag steps, creating a uniform sheet. Amazingly, as these 2D atomic layer islands merge, the boundary self-heals, maintaining the long-range phase coherence in atomic registry (> 120 nm, as shown in Figure 3c).

We next utilize scanning tunneling spectroscopy to study the electronic structure of the h-BeO for the first time. The differential conductance (dI/dV) spectroscopy, representing the local density of states (LDOS), is shown in Figure 5a with spectra spanning 5 orders of magnitude in dynamic range and displayed in logarithmic scale, acquired at a set bias of 4.1 V. With such a large dynamic range, we can assign the location of the CBM at 3.0 ± 0.2 eV and VBM at 3.0 ± 0.2 eV, corresponding to a bandgap of ~ 6 eV, similar to the value of previous theoretical calculations.[16, 25, 29-32] As discussed above, this explains why the layers resemble pristine Ag(111) at low sample bias (Figures 2b, 3b); within the bandgap, the h-BeO does not contribute to the LDOS. At a sample bias above the CBM, (Figs. 2a, 3c) tunneling from the BeO layer begins and the islands become visible. To further understand the interaction between the Ag(111) surface and the BeO, tunneling spectra (with a set point bias of 0.8 V, well within the BeO gap) was acquired on the clean Ag (black curve), and on the monolayer BeO surface (red curve), respectively (Figure 5b). Interestingly, the Ag surface state is preserved in the BeO region except for an upward shift of 0.12 ± 0.03 eV. This behavior closely resembles that of systems that are weakly interacting via vdW forces, for example hBN and Xe on Cu(111), which shift the Cu(111) surface state upward by ~ 0.1 eV.[34, 35] The preservation of the Ag (111) surface states is strong evidence that similarly, the h-BeO layer is weakly interacting with the Ag substrate.

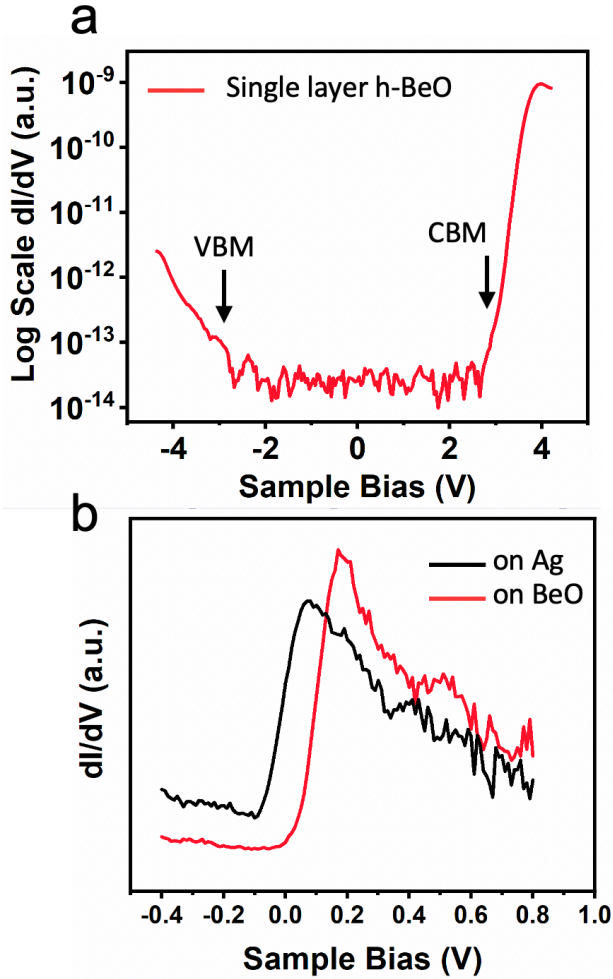

**Figure 5.** (a) Log scale dI/dV tunneling spectra acquired on the h-BeO layer at a set point bias voltage of 4.1 V, showing a bandgap of ~6 eV. (b) dI/dV tunneling spectra of the Ag surface states acquired on a single sheet of h-BeO (red curve) and on the Ag surface (black curve) at a set point bias of 0.8 V. Note that the Ag surface state is preserved with the addition of the h-BeO, with a slight upshift in energy.

As mentioned before, the observed h-BeO/Ag(111) moiré pattern displays a peculiar bias dependent contrast reversal, as shown in Figures 2a and 3c. This implies that the moiré patterns not only introduce a change in height corrugation, but also a spatial modulation of the electronic character. Here we adopt the terminology used in this field and call different regions in the moiré pattern "hole" and "wire" regions, as labeled in the inset of Figure 6a. Figure 6b is an STM image that illustrates the contrast reversal of the h-BeO that shows the hole region (white dashed line) changing from a protrusion to a depression when the bias was changed from $V_s = 3.7$ V (top) to 4.5 V (bottom). To investigate the contrast mechanism, we use Z-V spectroscopy to examine the field emission resonance (FER), as the sample bias of the first FER is a good approximation for the local work function. In this spectroscopic technique, the tunneling current is held constant while the tip-sample distance (Z) versus bias sweep is recorded. When the bias reaches the tunneling barrier resonance, a steep increase in Z occurs, therefore peaks in the derivative $(\partial Z/\partial V)_I$ denote the locations of the FER. Figure 6a shows the Z-V (top) spectra and their derivatives $(\partial Z/\partial V)_I$ (bottom) acquired at the hole region (red curve) and wire region (blue curve). We find that energy locations of FER in the hole region are consistently lower than those in the wire regions by about 0.08 eV, suggesting that the local work function is modulated across the moiré pattern. In addition, comparison of the h-BeO $(\partial Z/\partial V)I$ spectra with spectra obtained on an Ag(111) surface (black curve) shows the h-BeO layer significantly reduces the work function, explaining the darker contrast of the h-BeO regions at low bias as discussed previously (Figure 2b,c). A closer look at the work function modulation explains the variation of the moiré pattern contrast observed by STM (Figure 6b). In Figure 6c, we plot the tip height difference between the hole and wire regions directly deduced from the Z-V curves in Figure 6a and find that the positive values indicate the bias range where the hole regions appear as a protrusion, while at negative values it becomes a hole. We note that this work function behavior closely mirrors the trends observed for hBN on transitions metals,[35-40] providing further evidence that the interaction of the h-BeO and Ag(111) is similar to a vdW interface.

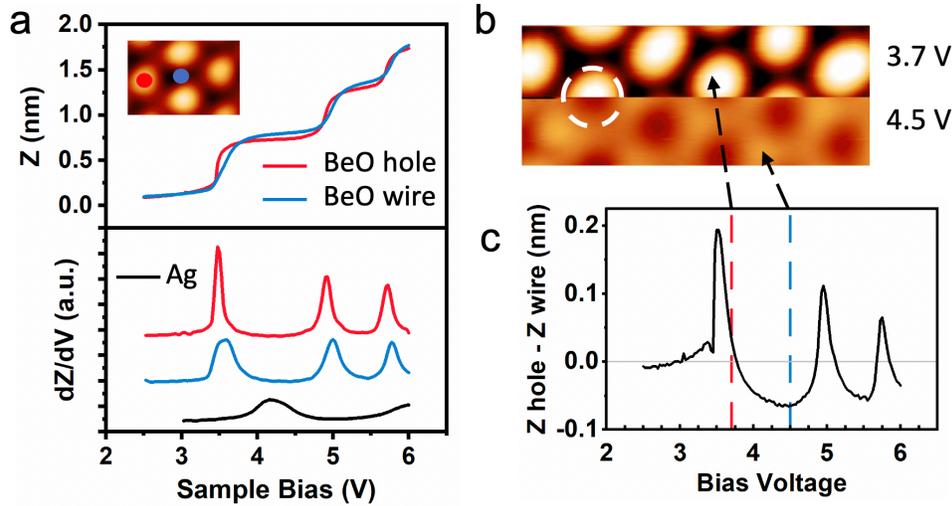

**Figure 6.** (a) The inset shows the different regions defined as the "hole" (red dot) and "wire" (blue dot) regions of the h-BeO moire pattern. The top graph shows vertical tip position (Z) as a function of the sample bias at constant current, and the bottom graph shows the corresponding dZ/dV spectra acquired at the hole and wire regions. The peaks indicate the location of the FER and show that the work function is modulated between the hole (red), wire (blue), and Ag (black). (b) STM image (area: 15 x 6 nm²) acquired at a sample bias of 3.7 V (top) and 4.5 V (bottom) showing a contrast reversal between the hole and wire regions. (c) The relative difference of the vertical tip height between the hole region and wire region with positive values indicating bias where the hole becomes bright, and negative values where the wire is bright.

To further understand the BeO/Ag interaction, we have performed first-principles calculations using a supercell consisting of 12×12 unit cells of monolayer honeycomb BeO on 11×11 unit cells of Ag(111), which gives rise to a 3.17-nm moiré pattern. The optimized structure is shown in Figure 7a, with regions marked by black, blue, and orange circles to represent the local stacking patterns of $Be_{fcc}O_{hcp}$, $Be_{top}O_{fcc}$, and $Be_{hcp}O_{top}$, respectively. This notation describes the locations of the Be and O atoms with respect to the surface sites on Ag(111); for example, $Be_{fcc}O_{hcp}$ corresponds to the configuration with Be at the fcc site and O at the hcp site. Due to the spatial variation of the stacking pattern, the interaction of the h-BeO layer with the Ag substrate varies within the moiré cell. As a result, the BeO-Ag interlayer distance fluctuates with a height corrugation from 3.02 to 3.47 Å which is a reasonable separation for a vdW interaction and agrees well with our measurements. The average height of each marked region is shown in Figure 7a, with the height of the black region set as the zero point. The orange region ($Be_{hcp}O_{top}$) with O on top of Ag has the strongest interaction and corresponds to the "hole" region seen in the STM scans. We further investigate the local work function variation along the diagonal of the supercell from left to right in Figure 7b. The work function values at black, blue, and orange regions are 4.26, 4.28, and 4.19 eV, respectively, which correlates to a modulation between the hole and wire region of roughly 70 – 90 meV and is consistent with the STM observation of 80 meV using barrier resonances as discussed above.

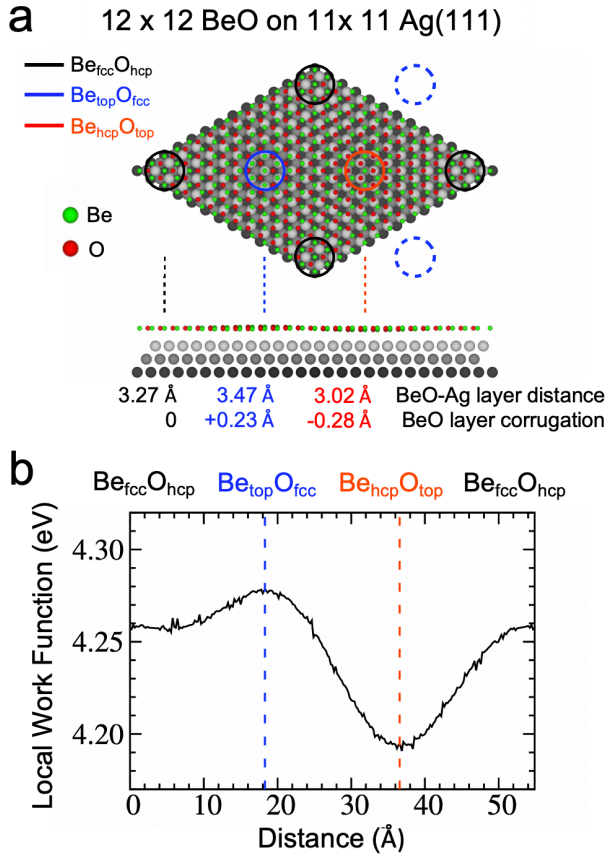

**Figure 7.** (a) Top and side views of a supercell containing 12x12 h-BeO/11x11 Ag(111) were used to simulate a 3.17-nm moiré pattern. The green and red spheres represent Be and O atoms, respectively. The 1st, 2nd, and 3rd layers of Ag atoms are illustrated by gray spheres with different darkness. High symmetry stacking regions of $Be_{fcc}O_{hcp}$, $Be_{top}O_{fcc}$, and $Be_{hcp}O_{top}$ are marked by black, blue, and orange circles, respectively. The dashed blue circles indicate the $Be_{top}O_{fcc}$ regions in the neighboring moiré cells. The BeO-Ag interlayer distance and the h-BeO layer corrugation are also shown at the bottom. (b) Calculated local work function along the diagonal of the supercell shown in (a).

CONCLUSION In summary, this work represents the compelling demonstration of the epitaxial monolayer growth of the graphitic phase of BeO by MBE, and also the direct characterization of the h-BeO atomic structure and electronic properties by STM/S. We find the h-BeO lattice constant of 2.65 ± 0.01 Å, and that the h-BeO surprisingly displays a nearly perfect atomic registry even across Ag steps like true vdW atomic layers. Previous studies of other wurtzite structure materials such as ZnO have shown that monolayers are observed at a local scale of a few tens of nm laterally, and exhibit multilayer, polycrystalline growth at higher coverage.[19-21] In contrast, our LEED and STM measurements suggest the astonishing formation of wafer scale coverage of a single atomic sheet of honeycomb BeO that maintains long-range phase of the atomic lattice. In our case, the growth appears to be self-terminated at one atomic layer, implying a catalytic role played by the underlying Ag (111) substrate. Yet another similarity to vdW materials is the preservation of the pristine Ag (111) surface electronic structure, indicating a weak interaction with the h-BeO atomic sheet, an assertion that is supported by our complimentary DFT calculations. We note that the size of our Si substrates is constrained by our sample holders, which are limited by the STM sample stage. Considering the h-BeO is directly grown on epitaxial Ag films on Si wafers, such a method may be translated to a larger growth format, providing a powerful scalable technology platform. The large insulating band gap of 6 eV and weak substrate interactions of the h-BeO make it an attractive candidate for multilayer heterostructures and as a template to support other 2D materials. We expect that our work demonstrates that insulating h-BeO is an outstanding addition to the library of 2D materials and will create future possibilities for bottom-up fabrication of 2D devices.

**Methods**

Sample growth and STM Characterization. Our experiments were performed in a UHV system that combines a homebuilt STM with two homebuilt MBE systems, one for growth of the Ag thin films, and one for BeO growth and LEED measurements. All of the chambers have base pressures below 1 x $10^{-10}$ Torr. The growth of the Ag(111) thin film substrate begins with preparing Si(111) wafers with 0.5 mm thickness by cutting them into 10 x 4 mm2 pieces, compatible with the STM sample holders. The sample is prepared following the standard procedure of flashing (1400 °C) and annealing (950 °C) in UHV to prepare an atomically clean surface with a well-defined 7x7 reconstruction. Growth of the Ag(111) thin films was carried out at room temperature with a deposition rate of 3 nm/min. The average thickness of the Ag films used in this study is 150 nm. The as-grown Ag film already has a well-defined (111) orientation, however, the surface is relatively rough. Annealing the film at 500 °C for one hour yields a much smoother surface with well-defined triangular terraces whose step edges are parallel to the <110> directions. See Ref. 41 for details. Next, a monolayer sheet of BeO was grown epitaxially on the epitaxial Ag(111) thin films. A Knudsen-cell with a BeO crucible purchased from Fermi Instruments (> 99% purity) was used as the evaporation source with a source temperature of 1200 °C. The Ag(111) substrate was maintained at 500 °C during the growth. Under these conditions, the growth rate is roughly ~ 0.07 ML per minute. After the desired coverage was reached, the shutter to the BeO source was closed, and the BeO sample was annealed at the growth temperature for 15 minutes. The chamber pressure during the growth reached 8 x $10^{-10}$ Torr.

STM measurements were performed at 77 K with a tungsten tip. The bias voltages were defined as the sample bias, and the topographic images were taken using constant current mode. The STM topography images were processed using Gwyddion software, and the STS was processed using the Origin software.

DFT Calculations. Our first-principles calculations were performed using the Vienna ab-initio simulation package (VASP)[42, 43] with the projector augmented wave method.[44, 45] The slab contains three layers of Ag atoms with the bottom two layers fixed and a vacuum region of about 16 Å. A plane-wave cut-off energy of 500 eV was used, and only the Gamma point was sampled for this large supercell. The optB86b functional including the van der Waals correction[46] was adopted for the structural relaxation. A dipole correction scheme was used to eliminate the long-range effect arising from the surface dipole between monolayer BeO and the Ag substrate. The atomic positions were fully relaxed until the force on each atom was smaller than 0.01 eV Å$^{-1}$. For the local work function calculation, we took the difference between the Fermi level and the electrostatic potential at the center of the vacuum above a specific region.

AUTHOR INFORMATION

**Corresponding Author**

*Email: shih@physics.utexas.edu

**Present Addresses**

‡ Hefei National Laboratory for Physical Sciences at the Microscale, University of Science and Technology of China (USTC), Hefei, Anhui 230026, China

**Author Contributions**

# M.H. and H.Z. each contributed equally to this work.

**Notes**

The authors declare no competing financial interest

ACKNOWLEDGMENT

This research was partially supported by the National Science Foundation through the Center for Dynamics and Control of Materials: an NSF MRSEC under Cooperative Agreement No. DMR-1720595. This research was also supported with grants from the Welch Foundation (F-1672), the US National Science Foundation (DMR-1808751, EFMA-1542747, EFMA-1542798), the US Airforce (FA2386-18-1-4097), and Academia Sinica (AS-TP-106-M07). We thank Professor Deji Akinwande for stimulating discussions on the technological implications of BeO.